\documentstyle[12pt,epsfig]{article}
\newcommand\as{\alpha_{\rm{S}}}

\def\ep{\epsilon}
\def\ee{$e^+e^-$}
\def\beq{\begin{equation}}
\def\eeq{\end{equation}}
\def\beeq{\begin{eqnarray}}
\def\eeeq{\end{eqnarray}}
\def\cm{{\cal M}}

\def\bom#1{{\mbox{\boldmath $#1$}}}
\def\to{\rightarrow}

\def\np#1#2#3{Nucl.\ Phys.\ B#1 (19#3) #2}
\def\pl#1#2#3{Phys.\ Lett.\ #1B (19#3) #2}
\def\pr#1#2#3{Phys.\ Rev.\ D #1 (19#3) #2}
\def\prep#1#2#3{Phys.\ Rep.\ #1 (19#3) #2}
\def\prl#1#2#3{Phys.\ Rev.\ Lett.\ #1 (19#3) #2}

\def\zp#1#2#3{Z.\ Phys.\ C#1 (19#3) #2}

\begin{document}

\begin{titlepage}
\renewcommand{\thefootnote}{\fnsymbol{footnote}}
\begin{flushright}
     CERN-TH/96-342 \\ hep-ph/9612236
\end{flushright}
\par \vspace{10mm}
\begin{center}
{\Large \bf
Jet cross sections \\
\vspace{2mm} 
at next-to-leading order\footnote
{Invited talk presented 
at the Cracow International Symposium on Radiative Corrections 
(CRAD96), Cracow, Poland, August 1996. To appear in the Proceedings.}
}
\end{center}
\par \vspace{2mm}
\begin{center}
{\bf Stefano Catani}\\

\vspace{3mm}

{I.N.F.N., Sezione di Firenze}\\
{and Dipartimento
di Fisica, Universit\`a di Firenze}\\
{Largo E. Fermi 2, I-50125 Florence, Italy}

\vspace{7mm}

{\bf Michael H. Seymour}\\

\vspace{3mm}

{Theory Division, CERN}\\
{CH-1211 Geneva 23, Switzerland}
\end{center}

\par \vspace{2mm}
\begin{center} {\large \bf Abstract} \end{center}
\begin{quote}
We briefly summarize theoretical methods for carrying out QCD
calculations to next-to-leading order in perturbation theory.
In particular, we describe a new general algorithm
that can be used for computing arbitrary jet cross sections in
arbitrary processes and can be straightforwardly implemented in
general-purpose Monte Carlo programs.  
\end{quote}
\vspace*{\fill}
\begin{flushleft}
     CERN-TH/96-342 \\ December 1996
\end{flushleft}
\end{titlepage}

\section{Motivations}
\vspace*{-2mm}

During the last fifteen years many efforts have been devoted to carry out
accurate QCD calculations to higher perturbative orders. These calculations
are motivated by three main reasons. 

First of all, the comparison between perturbative calculations and experimental
data allow one to perform precision tests of QCD in the strong-interaction
processes that involve a large transferred momentum $Q$ \cite{nigel,QCDrev}. 
These 
tests are essential for measuring the strong coupling $\as(Q)$ and its running 
\cite{as} as predicted by asymptotic freedom.
Perturbative QCD studies are also important to evaluate the background for new
physics signals. An outstanding example of that is the current 
investigation \cite{nigel} of the discrepancy
between the single-inclusive jet distribution at large $p_t$, as measured by CDF
\cite{cdfjet}, and the QCD predictions.
More recently, a renewed interest in 
perturbative calculations has been motivated by phenomenological and
theoretical models of non-perturbative phenomena (see \cite{renormalon}
and references therein). Using 
these models and having under
control the perturbative component, one can use
experimental data on high-energy cross sections to extract information on 
the underlying non-perturbative dynamics.

To these aims, calculations at the leading order (LO) of the perturbative
expansion in the QCD coupling $\as(Q)$ are insufficient. In fact, just because 
of its perturbative nature, the running of the QCD coupling can be hidden in
higher-order corrections. Thus at LO the value of $\as$ is essentially 
undetermined and 
a LO calculation
predicts only the order of magnitude of a given cross section and the rough
features of a certain observable. The accuracy of the perturbative
QCD expansion is instead controlled by the size
of the higher-order contributions. Any definite perturbative QCD prediction
requires (at least) a next-to-leading order (NLO) calculation.

In general, NLO calculations are highly non-trivial. The first bottleneck one 
encounters in producing new NLO calculations for a certain process is the 
evaluation of the relevant matrix elements. In recent years new techniques
\cite{oneloop}
have been developed to compute QCD Feynman diagrams and most of the one-loop
five-point amplitudes are now available \cite{5pam,g4pam}.
However, even when the 
process-dependent matrix elements are known, there are practical 
difficulties in setting up a straightforward calculational procedure.
The physical origin of 
these difficulties is in the necessity of factorizing the long- and 
short-distance components of the scattering processes
and is reflected in the perturbative expansion by the presence of divergences.
QCD theorems guarantee that these divergences eventually cancel in the
evaluation of physical cross sections but do not prevent their appearance
in intermediate steps. Since single intermediate expressions are usually 
divergent, the numerical implementation of NLO calculations forms a second 
bottleneck.

The main issue one has to face is thus the following. On one side many 
different NLO calculations (i.e. calculations for different observables) for a 
certain process and, possibly, for many processes are warranted. On the other 
side each calculation is very complicated (see also Sect.~\ref{nlo}).

In particular, it is very important to reduce the second
bottleneck by setting up efficient and simple {\em methods}
for computing {\em arbitrary} quantities in a single process.
It is even more important to have at our disposal simple {\em algorithms}
for computing arbitrary quantities in {\em arbitrary} processes. The goal is
a universal algorithm that, in principle, can be used to construct a
general-purpose Monte Carlo program (not a Monte Carlo event generator) for
carrying out
NLO QCD calculations. Conceptually, such an algorithm could be used in the same
manner as some universal Monte Carlo event generators (e.g. HERWIG \cite{HW}):
any time one wants to compute a new quantity or to vary the experimental cuts,
one simply modifies the `user routine' accordingly; any time one wants to study
a different process, one simply enters the corresponding matrix elements.

A new general algorithm of this type was recently presented \cite{paper}.
It is based on two key ingredients: the {\em subtraction method\/} for 
the numerical cancellation of the divergences among different contributions; 
and the {\em dipole factorization theorems} for the universal 
(process-independent) analytical treatment of individual divergent terms.  

In this contribution, after a brief summary of general methods,
we describe these two ingredients 
and show some numerical results for the
specific cases of jets in \ee\ annihilation and deep-inelastic 
lepton-hadron scattering (DIS).

\section{NLO QCD calculations}
\label{nlo}

The general structure of a QCD cross section in NLO is the following
\beq
\label{sig}
\sigma = \sigma^{LO} + \sigma^{NLO} \;.
\eeq
Here the LO cross section $\sigma^{LO}$ is obtained by integrating the fully
exclusive cross section $d\sigma^{B}$ in the Born approximation over the phase
space for the corresponding jet quantity. Let us suppose that this LO 
calculation involves $m$ partons with momenta $p_k$ $(k=1,...,m)$ in the final 
state. Thus, we write
\beq
\label{sLO}
\sigma^{LO} = \int_m d\sigma^{B} \;,
\eeq
where the Born-level cross section is:
\beq
\label{dsbf}
d\sigma^{B} = d\Phi^{(m)}(\{p_k\}) \; 
|{\cal M}_m(\{p_k\})|^2 \;F_J^{(m)}(\{p_k\}) \;\;,
\eeq
and $d\Phi^{(m)}$ and ${\cal M}_m$ respectively denote the full phase space 
and the tree-level QCD matrix element to produce $m$ final-state partons. These
are the factors that depend on the process.

The function $F_J^{(m)}$ defines the physical quantity that we want to compute,
possibly including the experimental cuts.
Note that this quantity has to be a jet observable, that is, it has to be 
infrared and collinear safe: its actual value
has to be independent of the number of soft and collinear particles in the
final state. Thus, we should have (we refer to \cite{paper} for a more detailed
formal definition)
\beq
\label{fjm}
F_J^{(m+1)} \to F_J^{(m)} \;\;,
\eeq
in any case where the $m+1$-parton configuration on the left-hand side is 
obtained from the $m$-parton configuration on the right-hand side
by adding a soft
parton or replacing a parton with a pair of collinear partons carrying the same
total momentum.

Efficient techniques, based on helicity amplitudes \cite{HA} and
colour subamplitude decomposition \cite{MP},
are available for calculating tree-level matrix elements. Thus
the evaluation of the LO cross section does not present any particular 
difficulty. Even if $\sigma^{LO}$  cannot be computed analytically (because
${\cal M}_m$ is too cumbersome or the phase-space cuts in $F_J^{(m)}$  are
very involved), one can straightforwardly use numerical integration techniques,
for instance, a Monte Carlo program where the function $F_J^{(m)}$ is
given as `user routine'.

At NLO one has to consider the exclusive cross section
$d\sigma^{R}$
with $m+1$ partons in the final state and the one-loop correction $d\sigma^{V}$
to the process with $m$ partons in the final state:
\beq
\label{sNLO}
\sigma^{NLO} 
= \int_{m+1} d\sigma^{R} + \int_{m} d\sigma^{V} \;.
\eeq
The exclusive cross sections $d\sigma^{R}$ and $d\sigma^{V}$
have the same structure as the Born-level cross section in Eq.~(\ref{dsbf}),
apart from the replacements $| {\cal M}_m |^2 \to | {\cal M}_{m+1} |^2$
and $| {\cal M}_m |^2 \to | {\cal M}_m |^2_{(1-loop)}$. Here
$| {\cal M}_m |^2_{(1-loop)}$ denotes the QCD amplitude to produce $m$ 
final-state partons evaluated in the one-loop approximation.

The calculation of the loop integral in $| {\cal M}_m |^2_{(1-loop)}$ leads
to {\em ultraviolet}, {\em soft} and {\em collinear} singularities. The
ultraviolet singularities can be handled in a simple way within the loop
corrections by carrying out the renormalization procedure. Thus we can assume 
that
the virtual cross section in Eq.~(\ref{sNLO}) is given in terms of the 
renormalized matrix element and the ultraviolet divergences have been removed.

Soft and collinear 
singularities instead lead to the main problem. These singularities do not 
cancel 
within the sole $d\sigma^V$ and are accompanied by analogous 
singularities arising from the integration of the real cross section 
$d\sigma^R$. In the case of jet
quantities, adding the real and virtual contribution, these singularities cancel 
and the physical NLO cross section in Eq.~(\ref{sNLO}) is finite. This
cancellation is guaranteed by the property in Eq.~(\ref{fjm}). However, the 
cancellation mechanism is not tri\-vial because it does not take place at the 
integrand level. 

The two integrals on the right-hand side of Eq.~(\ref{sNLO}) are separately 
divergent so that, before any
numerical calculation can be attempted, the separate pieces have to be
regularized. The most widely used regularization procedure (actually, the only
regularization procedure that is gauge invariant and Lorentz invariant 
to any order of the QCD perturbative expansion) is obtained  by means of 
analytic continuation in a number of space-time dimensions $d=4-2\epsilon$ 
different from four. Using dimensional regularization, the divergences 
(arising out of the integration) are replaced
by double (soft and collinear) poles $1/\ep^2$ and single (soft or collinear)
poles $1/\ep$.
Thus the real and virtual contributions should be calculated independently,
yielding equal-and-opposite poles in $\ep$. These poles have to be combined 
and, after having achieved their cancellation, the limit $\ep \to 0$ can be
safely carried out.

In principle this computation procedure does not pose any problems. In practice,
it is not so. On one side, analytic calculations are 
impossible for all but the simplest quantities because of the involved 
kinematics for multi-parton configurations and of the complicated phase-space 
cuts relative to the definition of the jet observable. On the other side,
the use of numerical methods is far from trivial because real and virtual 
contributions have to be integrated
{\em separately\/} over different phase-space regions and because 
of the analytic 
continuation in the arbitrary number $d$ of space-time dimensions.

The most efficient solution to this practical problem consists in using a hybrid 
analytical/numerical procedure: one must somehow simplify and extract the
singular parts of the cross section and treat them analytically;
the remainder is treated numerically, independently of the full complications 
of the jet quantity and of the process.

\subsection{General methods and algorithms}

There are, broadly speaking, two general methods for doing that: 
the {\em phase-space slicing\/} method and the {\em subtraction\/} method.
Both the slicing \cite{KL} and the
subtraction \cite{ERT} methods were first used in the context of NLO
calculations of three-jet cross sections in \ee\ annihilation. Then they
have been applied to other cross sections, adapting the method each time
to the particular process.
Only recently has it 
become clear that both methods are generalizable in a process-independent
manner. The key observation is that the singular parts of the QCD matrix
elements for real emission can be singled out in a general way by using
the factorization properties of soft and collinear radiation \cite{BCM}.
Owing to this universality, the two methods have led to general algorithms
for NLO QCD calculations. 

In the context of the phase-space slicing method, an algorithm has been 
developed for jet cross sections  
in lepton and hadron collisions \cite{GG,GGK}.
The generalization of this method to include fragmentation functions
and heavy flavours is considered in Refs.~\cite{BOO,GKL}.

As for the subtraction method, two approaches are available for setting up
general algorithms. The `residue approach' introduced in Ref.~\cite{KS}
has been further generalized in Refs.~\cite{Frix,MNR,NT}.
The dipole formalism \cite{CSlett} has been completely worked out in 
Ref.~\cite{paper}.

The advent of these algorithms has made feasible NLO QCD calculations for 
multi-jet cross sections.
Monte Carlo programs have been constructed for most of the
physical processes that involve four particles at LO. For five-particle
processes, the three-jet cross section in hadron collisions in the simplified
case of pure-gluon subprocesses is available \cite{pp3jet}, as is  
the four-jet cross section in electron-positron annihilation in the 
approximation of large number of colours \cite{4jleading};
the full QCD results are expected to appear soon.

We refer to Sect.~12.2 of Ref.~\cite{paper} for 
a discussion of the
comparison among different general methods for NLO calculations.
In the rest of this contribution
we describe the approach, based on the dipole formalism.

\section{The subtraction method}
\label{subm}

The general idea of the subtraction method is to use the identity
\beq
\label{sNLO1}
\sigma^{NLO} = \int_{m+1} \left[ d\sigma^{R} - d\sigma^{A}  \right] 
+  \int_{m+1} d\sigma^{A} +  \int_m d\sigma^{V} \;,
\eeq
which is obtained by subtracting and adding back the same quantity
$d\sigma^{A}$. The cross section contribution $d\sigma^{A}$ has to fulfil two 
main properties. 

$i)$ Firstly, it must be
a proper approximation of $d\sigma^{R}$ such as to have
the same {\em pointwise\/} singular behaviour (in $d$ dimensions) as
$d\sigma^{R}$ itself. Thus, $d\sigma^{A}$ acts as a {\em local\/} 
counterterm for $d\sigma^{R}$ and
one can safely perform the limit $\ep \to 0$ under the integral sign
in the first term on the right-hand side of Eq.~(\ref{sNLO1}). This defines
a cross section contribution $\sigma^{NLO\, \{m+1\}}$ with $m+1$-parton 
kinematics that can be integrated numerically in four dimensions:
\beq\label{NLOmp1}
\sigma^{NLO\,\{m+1\}} = \int_{m+1} \left[ \left( d\sigma^{R} \right)_{\ep=0}
- \left( d\sigma^{A} \right)_{\ep=0}  \;\right] .
\eeq

$ii)$ The second property of $d\sigma^{A}$ is its analytic integrability (in $d$
dimensions) over the one-parton subspace leading to the soft and collinear 
divergences. In this case, we can rewrite the last two terms on the right-hand
side of Eq.~(\ref{sNLO1}) as follows
\beq
\label{sNLOm}
\sigma^{NLO\,\{m\}} = \int_m 
\left[ d\sigma^{V} +  \int_1 d\sigma^{A} \right]_{\ep=0} \;\;. 
\eeq
Performing the analytic integration $\int_1 d\sigma^{A}$, one obtains $\ep$-pole
contributions that can be combined with those in $d\sigma^{V}$, thus
cancelling all the divergences. The remainder is finite in the limit $\ep \to 0$
and thus defines the integrand of a cross section contribution 
$\sigma^{NLO\,\{m\}}$ with $m$-parton kinematics that can be integrated 
numerically in four dimensions.

The final structure of the NLO calculation is as follows
\beq
\label{sNLO2}
\sigma^{NLO} = \sigma^{NLO\,\{m+1\}} + \sigma^{NLO\,\{m\}} \;\;,
\eeq
and can be easily implemented in a `partonic Monte Carlo' program, which 
generates
appropriately weighted events with $m+1$ and $m$ final-state partons.

Note that, using the subtraction method, no approximation is actually performed
in the evaluation of the NLO cross section. Rather than approximating the
cross section, the subtracted contribution $d\sigma^A$ defines a 
fake cross section
that has the same dynamical singularities as the real one and whose
kinematics are sufficiently simple to permit its analytic integration.

The real cross section contribution $d\sigma^{R}$ has the following
general structure
\beq\label{dsrf}
d\sigma^{R} = d\Phi^{(m+1)} 
\; |{\cal M}_{m+1}(\{p_k\})|^2 
\;F_J^{(m+1)}(\{p_k\}) \;, 
\eeq
where $d\Phi^{(m+1)}$ and $|{\cal M}_{m+1}|^2$ depend on the process
and $F_J^{(m+1)}$ depends on the quantity we want to compute.
Obviously, for any given $d\sigma^R$ one can try to construct a corresponding
$d\sigma^A$ by properly approximating $d\Phi^{(m+1)},$ $|{\cal M}_{m+1}|^2$
and $F_J^{(m+1)}$. It is less obvious that one can use the subtraction method
to compute {\em arbitrary} quantities in a given process, because one needs
a fake cross section $d\sigma^A$ that depends only on the process and, hence,
is independent of the actual definition of the jet function $F_J^{(m+1)}$. 
It is still less
obvious that one can use the subtraction method to construct
a universal algorithm for computing arbitrary quantities in {\em arbitrary}
processes. To this purpose the fake cross section $d\sigma^A$ also has to be  
somehow independent of ${\cal M}_{m+1}$.

Our method to achieve this generality is based on the dipole formalism.

\section{Dipole formalism and universal subtraction term}

\subsection{Soft and collinear limits}

The starting point of the dipole formalism are the soft and collinear
factorization theorems for the QCD matrix elements. According to these
theorems, the singular behaviour in $d$ dimensions of 
a generic tree-level matrix element $\cm_{m+1}(p_1,...,p_{m+1})$ with
$m+1$ final-state partons can be obtained
by means of factorized limiting formulae that, respectively in the soft 
(when the parton momentum $p_j$ vanishes) and collinear (when the parton momenta
$p_i$ and $p_j$ become parallel) regions,
have the following structure
\beq
\label{slimit}
|\cm_{m+1}(p_1,...,p_j,...,p_{m+1}) |^2 \to 
|\cm_{m}(p_1,...,p_{m+1}) |^2 \;
{\otimes}_c \;{\bom J}^2(p_j) \;\;,
\eeq
\beq
\label{climit}
|\cm_{m+1}(p_1,...,p_j,p_i,...,p_{m+1}) |^2 \to 
|\cm_{m}(p_1,...,p_j+p_i,...,p_{m+1}) |^2 \;
{\otimes}_h \; P_{ij} \;\;.
\eeq
The notation in Eqs.~(\ref{slimit},\ref{climit}) is symbolic (see  
Ref.~\cite{paper} for more details) but sufficient to recall their main
features.

The contributions $\cm_{m}$ on the right-hand sides are the
tree-level matrix elements to produce $m$ partons and are respectively 
obtained from the original $m+1$-parton matrix element by removing the soft
parton $p_j$ or combining the two collinear partons $p_j$ and $p_i$ into a 
single-parton momentum. 

The other contributions on the right-hand sides are responsible for the soft 
and collinear divergences. 
The factor ${\bom J}^2(p_j)$ in Eq.~(\ref{slimit}) is the eikonal current for 
the emission of the soft gluon $p_j$, and $P_{ij}$ is the Altarelli-Parisi 
splitting function. These factors are {\em universal}: they do not depend 
on the process but only on the momenta and quantum numbers of the QCD partons 
in $\cm_{m}$. In particular, ${\bom J}^2(p_j)$ depends on the colour charges
of the partons in $\cm_{m}$, and $P_{ij}$ depends on their helicities.
Because of these colour and helicity correlations (symbolically denoted
by $\otimes_c$ and $\otimes_h$), Eqs.~(\ref{slimit},\ref{climit}) are not real
factorized expressions. Moreover, there is another important reason, due to
kinematics, why 
Eqs.~(\ref{slimit},\ref{climit}) cannot be regarded as true factorization 
formulae but rather as limiting formulae. Indeed, the tree-level matrix elements
in Eqs.~(\ref{slimit},\ref{climit}) are unambiguously defined only when 
momentum conservation is fulfilled exactly.
Since, in general, the $m+1$- parton phase space does not factorize into an
$m$-parton times a single-parton phase space, the right-hand sides of
these equations are unequivocally defined only in the strict soft and collinear 
limits.

Owing to their universality, the limiting formulae (\ref{slimit},\ref{climit}) 
can be used to approximate
the matrix element $|{\cal M}_{m+1}|^2$ in Eq.~(\ref{dsrf})
and thus to find a fake cross section $d\sigma^A$ that matches the
real cross section $d\sigma^R$ in all the singular regions of phase space.
However, the implementation of Eqs.~(\ref{slimit},\ref{climit})
in the calculation of QCD cross sections requires a
careful treatment of momentum conservation away from the soft and
collinear limits. Care also has to be
taken to avoid double counting the soft and collinear divergences in
their overlapping region (e.g. when $p_j$ is both soft and collinear to $p_i$).
The use of the dipole factorization theorem introduced in Ref.~\cite{CSlett}
allows one to overcome these difficulties in a straightforward way. 
\vspace*{-4mm}
\subsection{Dipole formulae}

The dipole factorization formulae have
the following symbolic structure 
\beq
\label{Vsim}
|\cm_{m+1}(p_1,...,p_{m+1})|^2 = 
|\cm_{m}({\widetilde p}_1,...,{\widetilde p}_{m})|^2 
\otimes {\bom V}_{ij}
+ \dots \;\;.
\eeq
The dots on the right-hand side stand for contributions that are not singular 
when $p_i\cdot p_j \to 0$. 
The dipole splitting functions ${\bom V}_{ij}$ are universal 
(process-independent) singular factors that
depend on the momenta and quantum numbers of the $m$ partons in the tree-level
matrix element $|\cm_{m}|^2$. Colour and helicity correlations are denoted by
the symbol $\otimes$. The set ${\widetilde p}_1,...,{\widetilde p}_{m}$
of modified  momenta on the right-hand side of Eq.~(\ref{Vsim})
is defined starting from the original $m+1$ parton momenta in such a way that
the $m$ partons in $|\cm_{m}|^2$ are physical, that is, 
they are on-shell and energy-momentum conservation is
implemented exactly:
\beq
\label{kin}
{\widetilde p}_i^{\, 2} = 0 \;, \;\;\;
{\widetilde p}_1+...+{\widetilde p}_{m} = p_1+...+p_{m+1}\;\;.
\eeq
The detailed expressions for these parton momenta and for the dipole splitting
functions are given in Ref.~\cite{paper}.

Apart from the presence of colour and helicity correlations, Eq.~(\ref{Vsim})
can be considered as a true factorization formula because its left-hand and
right-hand sides live on the same phase-space manifold. Equation (\ref{kin})
indeed guarantees that exact kinematics are retained in the definition of the
$m$-parton configuration $\{{\widetilde p}_1,...,{\widetilde p}_{m}\}$. These
$m$ parton momenta depend on $p_i$ and $p_j$ in such a way that in the soft and
collinear regions the $m$-parton configuration become indistinguishable from
the original $m+1$-parton configuration. Correspondingly, the dipole splitting
function ${\bf V}_{ij}$ is defined in order to coincide with the eikonal 
current and with the Altarelli-Parisi splitting function respectively in the
soft and collinear limits. 

It follows that Eq.~(\ref{Vsim}) provides a {\em single\/} formula that
approximates the real matrix element $|\cm_{m+1}|^2$
for an arbitrary process, in {\em all\/} of its singular limits. These limits
are approached smoothly, thus avoiding double counting
of overlapping soft and collinear singularities. The exact implementation of 
momentum conservation makes possible this smooth transition and the 
extrapolation of the limiting formulae (\ref{slimit},\ref{climit}) 
away from the soft and collinear regions.

\subsection{Universal subtraction term}

These main features of the dipole formulae allow us to construct a universal 
subtraction term with the following form
\beq
\label{dsA}
d\sigma^{A} = d\Phi^{(m+1)} \;\sum_{ij} 
\;|{\cal M}_{m}(\{{\widetilde p}_k\})|^2 \otimes {\bom V}_{ij} 
\; F_J^{(m)}(\{{\widetilde p}_k\}) \;\;.
\eeq
Note that the only dependence on the jet observable is in the jet-defining 
function $F_J^{(m)}$ 
and the only dependence on the process is in the tree-level matrix element
$|{\cal M}_{m}|^2$. These are the same $m$-parton functions as enter in the 
calculation of the Born-level cross section of Eq.~(\ref{dsbf}). The only
other ingredients needed to construct $d\sigma^A$ are the dipole splitting 
functions, which are completely process-independent and given once and for all
\cite{paper}. This specifies the universal character of Eq.~(\ref{dsA}):
the fake cross section $d\sigma^A$ used for the NLO calculation is 
straightforwardly obtained in terms of the sole 
(process-dependent)
information that is necessary for the corresponding LO calculation. 

Having the subtraction term in the explicit form (\ref{dsA}), we can discuss how
it fulfils the properties $i)$ and $ii)$ listed in Sect.~\ref{subm}. As for the
property $i)$, we note that there are several dipole terms on the right-hand 
side of Eq.~(\ref{dsA}).
Each of them mimics one of the
$m+1$-parton configurations in $d\sigma^{R}$ that are kinematically degenerate
with a given $m$-parton state. Any time the $m+1$-parton state in $d\sigma^{R}$
approaches a soft and/or collinear region, there is a corresponding dipole
factor in $d\sigma^{A}$ that approaches the same region with exactly the
same probability as in $d\sigma^{R}$. The equality of the two probabilities 
directly follows from (\ref{dsA}) and from the limiting behaviour in 
Eqs.~(\ref{fjm},\ref{slimit},\ref{climit}) of the cross section factors 
on the right-hand side of Eq.~(\ref{dsrf}).
In this manner $d\sigma^{A}$ acts as a local counterterm for $d\sigma^{R}$.
Note, in particular, that the cancellation mechanism 
is completely independent of the actual
form of the jet-defining function and works for any jet observable (i.e.
for any quantity that fulfils Eq.~(\ref{fjm})).

As for the property $ii)$, we start by noting that $d\sigma^A$ (likewise 
$d\sigma^R$) depends on the $m+1$ parton momenta $p_1,...,p_{m+1}$.
However, having introduced the modified momenta ${\widetilde p}_1,...,
{\widetilde p}_{m}$, for each dipole term in Eq.~(\ref{dsA}) we can define
a {\em one-to-one} mapping 
\beq
\{p_1,...,p_{m+1}\} \leftrightarrow 
\{{\widetilde p}_1,...,{\widetilde p}_{m},p_i+p_j\} \;\;.
\eeq
The key feature of this mapping is that the $m$ modified momenta can be 
chosen in such a way that they obey {\em exact} phase-space factorization
as follows
\beq
\label{fac}
d\Phi^{(m+1)}(p_1,...,p_{m+1}) = 
d\Phi^{(m)}({\widetilde p}_1,...,{\widetilde p}_{m}) 
\; d\varphi_{(\{{\widetilde p}_k\})}(p_i+p_j) \;,
\eeq
where $d\varphi$ is a single-particle subspace that, for fixed 
${\widetilde p}_1,...,{\widetilde p}_{m}$, depends only on the dipole momenta
$p_i$ and $p_j$ \cite{paper}. Owing to the exact phase-space factorization and 
to the fact that the fake cross section in Eq.~(\ref{dsA}) is proportional to
the jet quantity calculated from the modified $m$-parton configuration,
the integration of the singular dipole contributions can be completely 
factorized
(modulo colour and helicity correlations) with respect to a term that exactly
reproduces the Born-level cross section:
\beeq
\label{dsA1}
\!\!\!\!\!\!\!\int_{m+1} d\sigma^{A} &\!\!=\!\!& \int_m 
d\Phi^{(m)}(\{{\widetilde p}_k\}) 
\;|{\cal M}_{m}(\{{\widetilde p}_k\})|^2 \; F_J^{(m)}(\{{\widetilde p}_k\})
\nonumber \\
&\!\!\otimes\!\!& \;\sum_{ij} \int_1 
\;d\varphi_{(\{{\widetilde p}_k\})}(p_i+p_j)
\;{\bom V}_{ij} 
= \int_m  \;d\sigma^{B}  \otimes  
{\bom I}(\{{\widetilde p}_k\}) \;.
\eeeq
The last factor on the right-hand side of Eq.~(\ref{dsA1}) is defined by
\beq
\label{Ifac}
{\bom I}(\{{\widetilde p}_k\}) \equiv  
\sum_{ij} \int_1 \;d\varphi_{(\{{\widetilde p}_k\})}(p_i+p_j)
\;{\bom V}_{ij}, \;\;
\eeq
and contains all the soft and collinear singularities
that are necessary to compensate those in the virtual cross section $d\sigma^V$.
Owing to the convenient definition of the dipole splitting function 
${\bom V}_{ij}$, it is possible to carry out analytically the integration 
in Eq.~(\ref{Ifac}) over the dipole phase space in $d$
dimensions. This leads to an explicit and universal expression \cite{paper}
for the factor ${\bom I}$, whose $\ep$-poles cancel those in the one-loop
matrix element.

\section{Final results and numerical implementation}

The discussion in the previous section shows that, by using the subtraction
method and the dipole formulae, one can extract and treat analytically
the singular parts of any NLO cross section in a way that is independent of the
exact details of the observable and of the process.
This leaves a remainder that depends on the full complications of the jet 
quantity, but which is finite so that it can be treated either numerically or 
analytically (whenever possible).

In general, the use of numerical integration
techniques (typically, Monte Carlo methods) is certainly more convenient. 
First of all, the numerical approach allows one to calculate any number and
any type of observable simultaneously by simply histogramming the appropriate
quantities, rather than having to make a separate analytic calculation for each
observable. Furthermore, using the numerical approach, it is easy to
implement different experimental conditions, for example detector acceptances
and experimental cuts. 

In order to summarize the final results of our algorithm and to describe their
numerical implementation, 
we start by recalling how the LO cross section in Eq.~(\ref{sLO}) is 
evaluated by using a Monte Carlo program. One first generates an 
$m$-parton event in the phase-space region $d\Phi^{(m)}$
and gives it the weight $|\cm_{m}|^2$. Then this weighted event is analysed
by a user routine according to the actual definition of
the phase-space function $F^{(m)}_J$ and inserted into a corresponding 
histogram bin.

Following the decomposition in Eq.~(\ref{sNLO2}), the NLO cross section is 
obtained by adding two contributions (which are not necessarily
positive definite) with $m$-parton  (as in the
LO calculation) and $m+1$-parton kinematics, respectively. Unlike the original
real and virtual contributions, these two terms are separately finite and can
directly be integrated in four space-time dimensions. 

\subsection{The term with $m$-parton kinematics}

The first contribution is obtained by inserting Eq.~(\ref{dsA1}) 
into Eq.~(\ref{sNLOm}) and 
can be written as follows
\beq
\label{sNLOMe}
\sigma^{NLO\,\{m\}} =
\int_{m} d\Phi^{(m)} \;F_J^{(m)}(\{p_k\})  \;{\cal F}_m(\{p_k\}) \;\;,
\eeq
where the {\em master function} ${\cal F}_m(\{p_k\})$ is explicitly given by
\beq
\label{master}
{\cal F}_m(\{p_k\}) = \left\{ \; | \cm_{m}(\{p_k\})|^2_{(1-loop)}   
+  \frac{}{} | \cm_{m}(\{p_k\})|^2 \otimes {\bom I}(\{p_k\}) 
\; \right\}_{\ep=0} \;.
\eeq
The first term in the curly bracket is the one-loop {\em renormalized\/} matrix
element for producing $m$ final-state partons.
The second term is obtained by combining the 
tree-level matrix element to produce $m$ partons and the universal
factor ${\bf I}$ in Eq.~(\ref{Ifac}). These two terms are defined in 
$d=4-2\ep$ dimensions. Owing to the progress made in recent years in the 
analytical techniques for evaluating one-loop amplitudes \cite{oneloop}, many 
of them have been calculated. The explicit expression of the
universal factor ${\bf I}$ is provided by our algorithm. Thus, one has to
carry out the expansion in $\ep$-poles of the two terms in the curly bracket,
cancel analytically (by trivial addition) the poles and perform the limit
$\ep \to 0$. This simple algebraic manipulation is sufficient to construct
an effective $m$-parton weight, the master function $\cal F$,
that is finite. As a result,  
Eq.~(\ref{sNLOMe}) can be handled by the Monte Carlo program 
exactly in the same way as the LO cross section.

Note that the two terms on the right-hand side of Eq.~(\ref{master}) separately 
depend on the regularization prescription of the soft and collinear divergences,
namely dimensional regularization. Since different versions of 
dimensional regularization can be used to compute the one-loop matrix element,
the second term in the curly bracket has to be evaluated accordingly. 
Alternatively, one can fix the latter and use the transition rules derived in 
Ref.~\cite{KST2to2} to relate the one-loop amplitudes in different 
dimensional-regularization schemes.

The necessity to consistently regularize the separately divergent components 
of the cross section is a common feature of any NLO calculation, independently 
of the method that is actually used in the computation. Failure in the 
consistent implementation of the regularization procedure leads to
violation of unitarity and, ultimately, to an incorrect (although possibly 
finite) final result. The dipole formalism is extremely efficient to guarantee
unitarity because all the divergences are isolated in the right-hand side of
Eq.~(\ref{master}). As explained in Ref.~\cite{uni}, for any regularization 
prescription that is unambiguously defined at the level of one-loop matrix 
elements, one can compute in a simple and consistent way the universal factor
$\bom I$ that provides the finite and unitary master function $\cal F$.

\subsection{The term with $m+1$-parton kinematics}

The NLO contribution with $m+1$-parton kinematics, which is obtained by 
subtracting 
the fake cross section in Eq.~(\ref{dsA}) from the real cross section in 
Eq.~(\ref{dsrf}), has the following explicit expression:
\beeq
\label{sNLOm1}
\!\!\!\!\!\!\!\!\!\!\!\!&&\!\!\!\!\!\!\!\!\!\!\sigma^{NLO\,\{m+1\}} = 
\int_{m+1} d\Phi^{(m+1)} \!\! \\
&&\!\!\!\!\!\!\!\!\!\!\cdot \left\{ 
| \cm_{m+1}(\{p_k\})|^2 \;F_J^{(m+1)}(\{p_k\}) 
-  \sum_{ij} | \cm_{m}(\{{\widetilde p}_k\})|^2 \otimes
{\bom V}_{ij} F_J^{(m)}(\{{\widetilde p}_k\}) \right\} . \nonumber 
\eeeq
The terms in the curly bracket define an effective matrix element
that is integrable in four space-time dimensions. It follows that the NLO
matrix element $\cm_{m+1}$, with $m+1$ final-state partons, can be directly 
evaluated in $d=4$ dimensions, thus leading to an extreme simplification of
the Lorentz algebra. Knowing the tree-level matrix elements and the dipole
splitting functions, the Monte Carlo integration of Eq.~(\ref{sNLOm1}) is
straightforward. One simply generates an $m+1$-parton configuration
and uses it to define an event with positive weight $+|\cm_{m+1}|^2$ and
several counter-events, each of them with the negative weight 
$-|\cm_{m}|^2 \otimes {\bf V}_{ij}$. Then these
event and counter-events are analysed by the user routine.
The role of the two different jet functions $F_J^{(m+1)}$ and $F_J^{(m)}$
is that of binning the weighted event and counter-events into different bins of 
the jet observable. Any time that the generated $m+1$-parton
configuration approaches a singular region, the event and one counter-event
fall into the same bin and the cancellation of the large  positive
and negative weights takes place.

\section{Monte Carlo programs}

Generalizing the procedure for constructing NLO Monte Carlo programs for
arbitrary quantities has several advantages.  These are principally due to
the reduction in the number and complexity of ingredients that have to
be calculated for each new process, and because the $d$-dimensional
integrals only need be done once and can be easily checked independently,
rather than being buried inside a specific calculation.

Using the general algorithm described in this contribution, we have already
constructed two Monte Carlo programs 
(they can be obtained from
{\verb+http://surya11.cern.ch/users/seymour/nlo/+}), 
EVENT2 and DISENT.

\begin{figure}
\centerline{\epsfig{figure=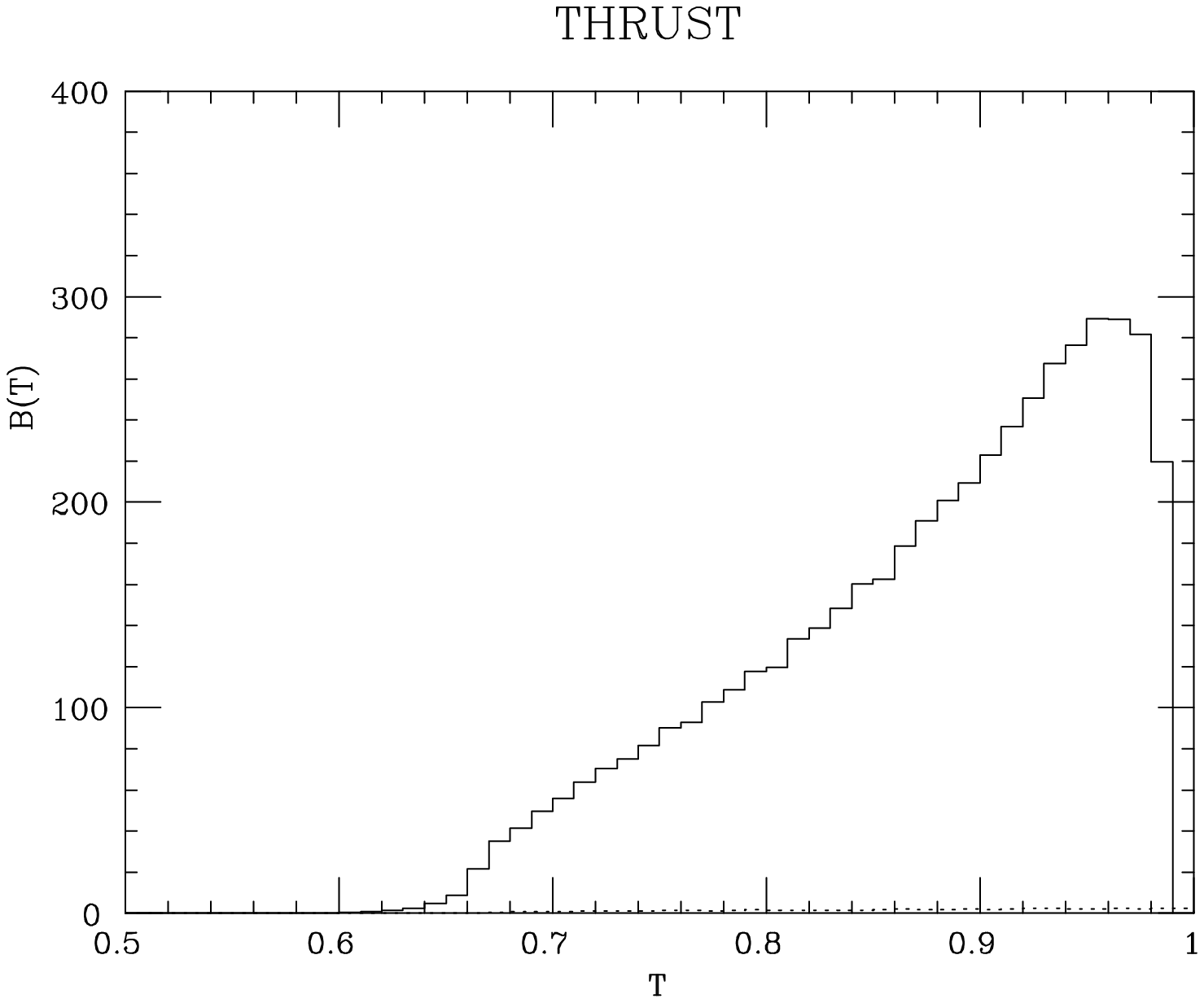,height=7cm}}
\vspace*{7mm}
\centerline{\epsfig{figure=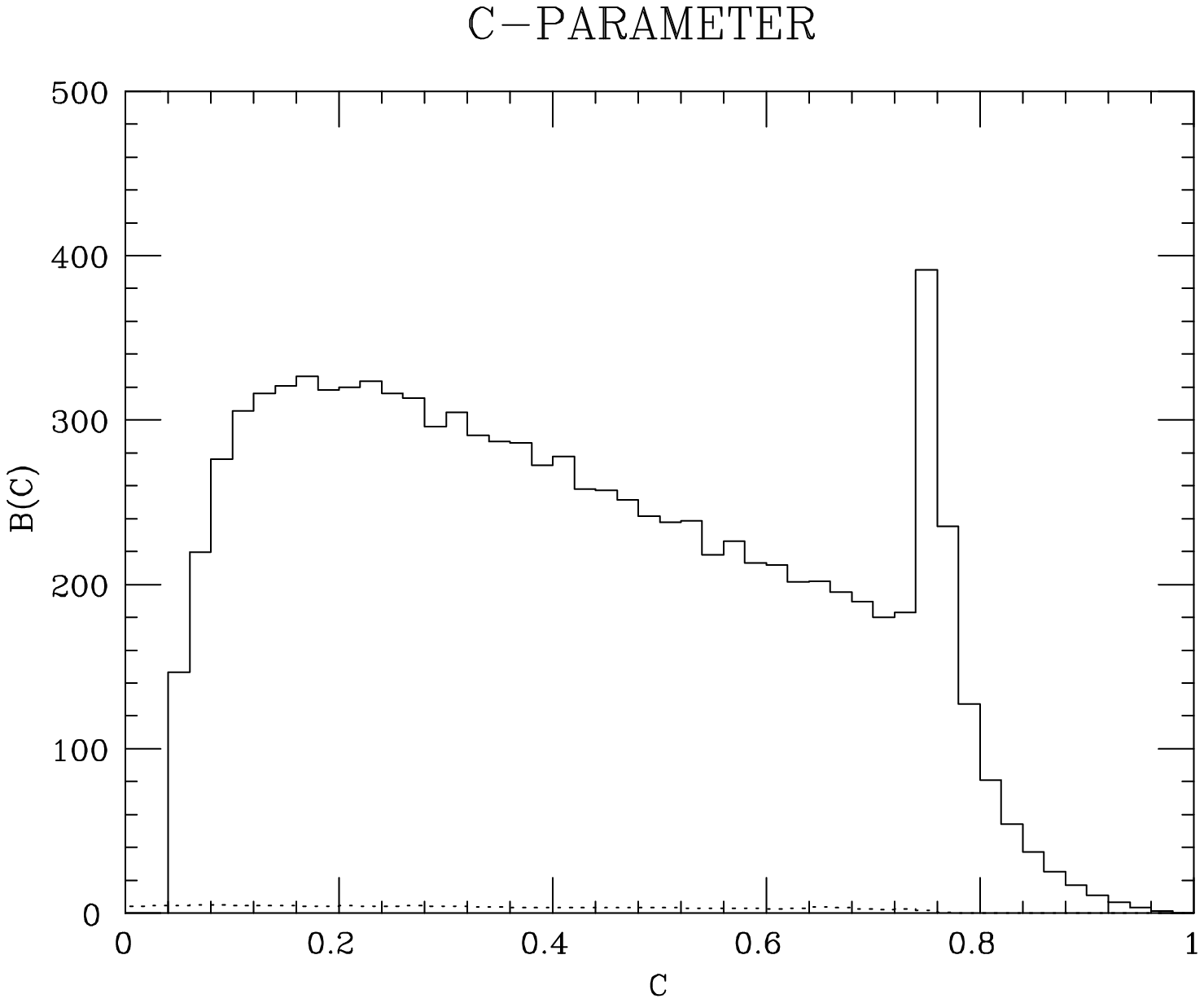,height=7cm}}
  \caption[]{Coefficient of $(\as/2\pi)^2$ for the thrust and $C$-parameter
    distributions. The dotted histograms show the size of the statistical
    errors.}
  \label{thefig}
% The bounding box for the figures is %%BoundingBox: 100 240 525 580
\end{figure}

EVENT2 \cite{CSlett} computes three-jet observables
in \ee\ annihilation. In the case of un-oriented three-jet events,
this program is comparable and in agreement with the program EVENT \cite{KN},
which is based on the subtraction procedure of Ref.~\cite{ERT} and has been
used for most of the QCD analyses at LEP and SLC \cite{QCDrev,as}.
As an example we show the NLO coefficients for the thrust and $C$-parameter
distributions in Fig.~\ref{thefig}. We find that, in general, the numerical 
convergence of EVENT2 is similar to the program of
Ref.~\cite{KN}, except close to the two-jet region in which ours becomes
progressively better. In the case of oriented events \cite{lep2}, EVENT2 
should be compared with 
a corresponding program, EERAD \cite{GG}, based on the phase-space slicing
method.

\begin{figure}
\centerline{\epsfig{figure=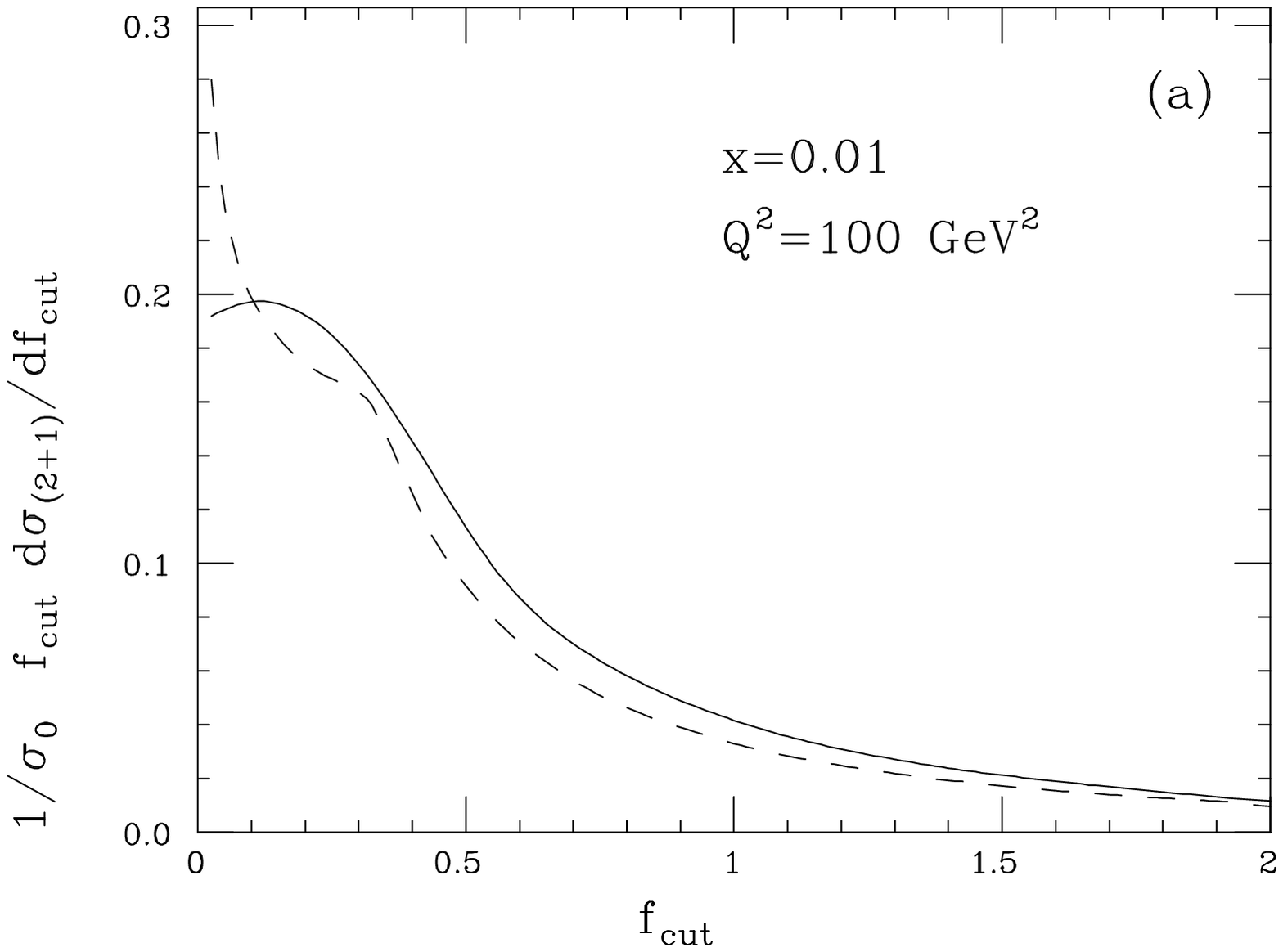,height=6cm}}
\vspace*{3mm}
\centerline{\epsfig{figure=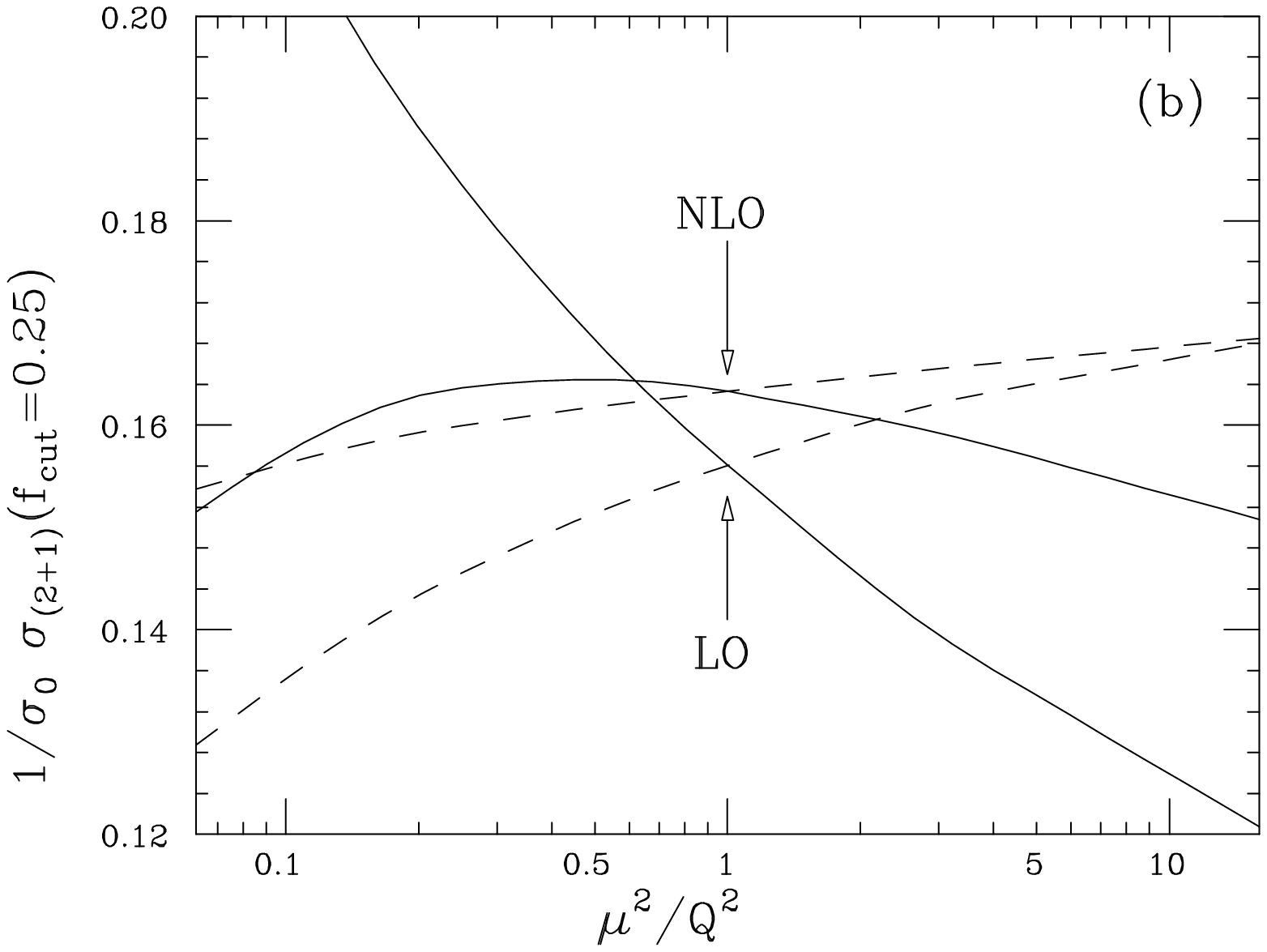,height=6cm}}
\caption[]{ Jet cross sections in $ep$ collisions at HERA energies 
            (${\sqrt s}= 300~{\rm GeV}$).
           (a) The distribution of resolution parameter $f_{cut}$ at which
                DIS events are resolved into $(2+1)$ jets according to the
                $k_\perp$ jet algorithm.  Curves are LO (dashed) and NLO
                (solid) using factorization and renormalization scales
                equal to $Q^2$, and the ${\rm MRS D'_{-}}$ distribution 
                functions. Both curves are normalized to the LO cross section.
           (b) The rate of events with exactly $(2+1)$ jets at
                $f_{cut}=0.25$ with variation of renormalization
                (solid) and factorization (dashed) scales.  Normalization
                is again the LO cross section with fixed factorization
                scale.
\label{fig}}
\end{figure}

DISENT \cite{paper,hera} is a NLO program for $2+1$-jet observables in DIS.
The program uses the matrix
elements evaluated by the Leiden group \cite{leiden}. 
In Fig.~\ref{fig}a
we show as an example the differential jet rate as a function of jet
resolution parameter $f_{cut}$,
using the $k_\perp$ jet algorithm~\cite{ktalg} at HERA energies~\cite{zar}.  
We see that the NLO 
corrections are
generally small and positive, except at very small
$f_{cut}$
(where large logarithmic terms, $-\alpha_s\log^2f_{cut}$, arise at each
higher order).
In Fig.~\ref{fig}b, we show the variation
of the  
jet rate at a fixed $f_{cut}$ with factorization and
renormalization scales. The scale dependence is considerably smaller at NLO.
DISENT can be compared with the Monte 
Carlo MEPJET \cite{Mirkes} that uses the phase-space slicing algorithm
of Ref.~\cite{GGK}.
 
The results of the algorithm based on the dipole formalism have also
been implemented in a program \cite{CC10} for the calculation of 
NLO QCD corrections to four-fermion final states in \ee\ annihilation.

\section{Summary and outlook}

The calculation of jet cross sections in perturbative QCD requires the 
integration of multiparton matrix elements over complicated phase-space regions
that depend on the actual definition of the jet observables and on the
experimental cuts. In general, these phase-space integrations can be carried out
only by using numerical methods. Beyond LO, however, numerical techniques
cannot straightforwardly be applied because real-emission contributions and 
virtual contributions are separately divergent. These divergences have to be 
first regularized, then evaluated analytically, combined together and cancelled
before any numerical calculation can be attempted.  

General methods are now available to 
overcome all the analytical difficulties related to the 
treatment of soft and collinear divergences in NLO calculations.
In this contribution we have mainly described one of these general formalisms,
which has been used to set up an explicit algorithm to compute NLO jet 
cross sections.

The algorithm combines  
the subtraction method and the dipole formulae to
carry out all the analytical work that is necessary to evaluate and
cancel the singularities. The final output of the algorithm is given in terms
of effective matrix elements that can be automatically constructed starting
from the original (process-dependent) matrix elements and universal 
(process-independent) dipole factors. The effective matrix elements
can be integrated numerically or analytically (whenever possible)
over the available phase space in four dimensions to compute the actual value 
of the NLO cross section. If the numerical approach is chosen, Monte Carlo 
integration techniques can be easily implemented to provide a general-purpose 
Monte Carlo program for carrying out NLO QCD calculations in any given process.

The simplified discussion of the algorithm presented in this contribution
directly applies to processes, like \ee\ $\to n$ jets, in which there are 
neither
initial-state hadrons nor identified hadrons in the final state. However,
the formalism and the algorithm are completely general in the sense that they
apply to {\em any} jet observable in a given scattering process as well as 
to {\em any} hard-scattering process.
Full details and explicit results for lepton-hadron and hadron-hadron collisions
and for fragmentation processes are given in Ref.~\cite{paper}.

At present, 
next-\-to-\-next-\-to-\-leading order 
(NNLO) 
QCD calculations 
are feasible
only for some fully inclusive quantities \cite{NNLO}.
In these cases 
one considers all possible final states and integrates
the QCD matrix elements over the whole final-state phase space.
Thus one can add real and virtual contributions before performing the relevant
momentum integrations in such a way that only ultra\-violet singularities appear
at the intermediate steps of the calculation. In the case of less inclusive
jet observables, one cannot take advantage of the cancellation of soft
and collinear divergences at the integrand level and, at present, no  
systematic method is available to handle these divergences at NNLO.
Even once the necessary two-loop matrix elements for several processes
are calculated, 
the amount of work needed to
provide a numerical implementation will be enormous.  
The main features of the dipole formalism, which permit a universal 
treatment of soft and collinear singularities at NLO, seem particularly suited
to set up a general method for carrying out NNLO QCD calculations.

\vskip 5mm
\noindent{\bf Acknowledgements.} This research is supported in part by
EEC Programme `Human Capital and Mobility', Network `Physics at High Energy 
Colliders', contract CHRX-CT93-0357 (DG 12 COMA).
We would like to thank Staszek Jadach and the Local Organizing Committee
for the succesful organization of this Symposium.

\end{document}